\documentclass[12pt,onecolumn] {article} 
\usepackage{amsmath}
\usepackage{amssymb}
\usepackage{latexsym}
\usepackage{graphicx}

\begin{document}

\begin{center}
\baselineskip=24pt

{\Large \bf The expected background spectrum in NaI dark matter detectors and the DAMA result}

\baselineskip=18pt

\vspace{0.3cm}
V.~A.~Kudryavtsev \footnote{Corresponding author; address: 
Department of Physics and Astronomy, University of Sheffield,
Sheffield S3 7RH, UK, e-mail: v.kudryavtsev@sheffield.ac.uk},
M.~Robinson,
N.~J.~C.~Spooner

{\it Department of Physics and Astronomy, University of Sheffield, Sheffield, S3 7RH, UK}\\

\vspace{0.3cm}
\begin{abstract}
Detailed Monte Carlo simulations of the expected radioactive background 
rates and spectra in NaI
crystals are presented. The obtained spectra are then compared to those
measured in the DAMA/NaI and DAMA/LIBRA experiments.
The simulations can be made consistent
with the measured DAMA spectrum only by assuming higher than reported 
concentrations of some isotopes and even so leave very little room for the dark matter signal.
We conclude that any interpretation of the annual modulation of the event rate
observed by DAMA as a dark matter signal, should
include full consideration of the background spectrum. 
This would significantly restrict the range
of dark matter models capable of explaining the modulation effect.
\end{abstract}

\end{center}

\vspace{1.0cm}

\vspace{1.0cm}

\pagebreak

\section {Introduction}
\label{intro}
The DAMA group has reported a possible signal from dark matter particles. The first
evidence came from the DAMA/NaI experiment (100 kg of NaI), published in 
Refs. \cite{dama1,dama2}. The most recent papers \cite{dama3,dama4} 
based on DAMA/LIBRA measurements
using 250 kg of NaI combined with the DAMA/NaI results presented improved statistical significance
of the signal and evaluation of possible background sources and systematic uncertainties.
The positive identification of the effect from dark matter particles is based on observation
of the annual
modulation of the event rate at low energies 
with period (one year) and phase consistent with
those expected from dark matter halo models \cite{halo1,halo2}. 

Unlike many other experiments, DAMA/NaI and DAMA/LIBRA did not use pulse
shape discrimination between electron and nuclear recoils to eliminate the background
from gamma-rays for their high statistics exposure, 
relying only on the annual modulation of the total event rate
and energy spectrum. In the presence of a non-zero signal, the event rate 
in any particular energy bin can be described by the equation:

\begin{equation}
R(E,t) = b(E) + S_{0}(E) + S_{m}(E) \cos \left( \omega (t - t_0) \right),
\label{eq:1}
\end{equation}

where $R(E,t)$ is the time-dependent event rate in a particular energy bin, 
$b(E)$ is the time-independent rate of
background events, $S_{0}(E)$ is the time-independent
non-modulated part of the dark matter signal
(average event rate),
$S_{m}(E)$ is the amplitude of the modulated part of the dark matter signal, 
$\omega$ is the frequency
of modulation ($\omega = 2 \pi / T$, where  $T= 1$ year) and $t_0$ is the time 
in years for the maximum of the signal.
A similar equation has been used in the analysis of the DAMA data \cite{dama2,dama3}
but the time-independent terms $b$ and $S_0$ of the equation were combined
together. Such an approach, however, neglects the correlation between the
modulated and non-modulated parts of the signal.
Interpretation of the DAMA signal in terms of dark matter models 
carried out by the DAMA group 
\cite{dama2,dama3} and in many other papers (see, for instance \cite{fs08,sggf08,cpw08} 
for recent analyses) were limited to fits 
to the modulated signal and, in some cases, to reconstruction of 
the average energy spectrum of the signal $S_{0}(E)$.

According to Eq. (\ref{eq:1}), the total average rate of events in each energy bin
is the sum of the background rate $b(E)$ and the average rate of events
from dark matter particles $S_{0}(E)$, the third term being cancelled out when
averaged over a period of a year or several years. This means that if the 
background spectrum were known we would be able to check whether
the sum of the reconstructed spectrum of the signal ($S_{0}(E)$) and 
background ($b(E)$) matches
the measured distribution. It turns out, however, that most efforts so far 
were directed to calculation of the signal (mainly $S_{m}(E)$) 
in various models helping with interpretation of
the measured annual modulation effect, but totally neglecting the 
background contribution to the
event spectrum and the information about the dark matter models that can
be extracted from the measured spectrum.
The background event rate and spectrum can be simulated with a reasonable 
degree of accuracy
if the abundances of radioactive isotopes in detector materials are known.
The DAMA Collaboration should have included the evaluation of this spectrum in their results
to support their claim of dark matter discovery.

Limiting the data analysis to the modulated part of the spectrum only, and ignoring 
information from the non-modulated component of the signal and background, 
severely weakens the conclusions that may be derived from the experiment.
Self consistent analysis of the experimental data should include all components of the
measured spectrum: background, modulated and non-modulated parts of the
signal. In this way no information is missed and a proper fit of the measured spectrum to
the sum of predicted distributions is possible. In practice, accurate calculation of 
the background spectrum in terms of the absolute event rate (events/kg/day/keV)
can only be done with high accuracy if concentrations of radioactive isotopes in all materials
in and around the detector are precisely known. However, any particular source
of background radiation has a characteristic spectrum and a combination of spectra
from different sources with certain weights should match observations
with or without signal. Hence the difference between the measured and 
background spectra should be equal to the spectrum of dark matter events.
Analysing the measured spectrum and simulated background 
would allow us to constrain the dark matter signal.

We have used our expertise in running NaI(Tl) dark matter experiment
NAIAD \cite{naiad} and in modelling radioactive background in different
types of detectors \cite{carson1,carson2,vito} to simulate gamma-ray
production, transport and detection in the DAMA/LIBRA experiment.
In this paper we present detailed Monte Carlo simulations of the background spectrum
in a modelled setup of the DAMA/LIBRA NaI(Tl) detector. We then
discuss the implications that these results may have on interpretation
of the signal claimed by the DAMA group. 
We demonstrate that the energy spectrum measured in the DAMA/LIBRA experiment
can hardly include any dark matter signal by (i) subtracting some possible WIMP signals
from the measured spectrum and comparing this to the simulated background and
(ii) subtracting the simulated background spectrum from the measured one and showing
that the difference spectrum has a minimum at low energies which would not be a usual
feature for dark matter signals. We also show that some inconsistencies exist
between low-energy and high-energy parts of the measured DAMA/LIBRA spectrum.

A possible problem with
the background spectrum in the DAMA/NaI experiment in the presence
of dark matter signal was pointed out also 
in Ref. \cite{gilles}. Several recent analyses (see, for instance, \cite{fs08,cpw08})
used the measured spectrum to constrain
theoretical models but did not take into account the spectrum of
background events that should contribute to the observed rate.

\section{Modelling of the background radiation}
\label{modelling}

To model the DAMA/LIBRA background we 
have considered three locations of radioactive sources:
(i) external sources producing gamma-rays and other particles 
outside the NaI crystals; ii) internal sources of radiation evenly distributed
within the volume of the crystal; and iii) surface sources with
radioactive isotopes concentrated within the thin surface layer 
of the crystal (assumed here to be 50 $\mu$m thick).

\subsection{External background sources}

The setup of the DAMA/LIBRA experiment 
comprising 25 NaI(Tl) crystals, $\sim$10 kg each, was
modelled using the published geometry \cite{dama3,dama4}. 
For case (i) the source of background radiation was assumed to be in
the 100 g envelopes of the photomultiplier tubes (PMTs) attached to the light guides 
connected to the
two flat surfaces of the crystals. The exact position of the radiation source (PMTs, light guides
or similar) is not important
since only the high-energy gamma-rays can reach the crystal and the spectrum
of events at low energies (below 30 keV) is fully determined by Compton 
electrons. 
The PMT envelopes were populated with radioactive isotopes of 
$^{238}$U, $^{232}$Th and $^{40}$K. The isotopes were 
allowed to decay in our Monte Carlo code, based
on the GEANT4 toolkit (version 9.2) \cite{geant4} . All particles were generated according to 
GEANT4 library. All isotopes in the decay chains of $^{238}$U and $^{232}$Th
decayed in our simulations in secular equilibrium (the spectra
of gamma-rays from uranium and thorium decay chains in secular equilibrium 
have also been obtained in Ref. \cite{pandola}).
Obviously, only high-energy gamma-rays can reach NaI from large distances, with
X-rays, electrons and alphas being absorbed in the surrounding materials.

We have also considered $^{60}$Co in copper around the detectors and in a shield,
as a possible source of background events. Simulations of $^{60}$Co induced events
have been carried out in the same way as for other isotopes.

\begin{figure}[htb]
   \includegraphics[width=15.cm]{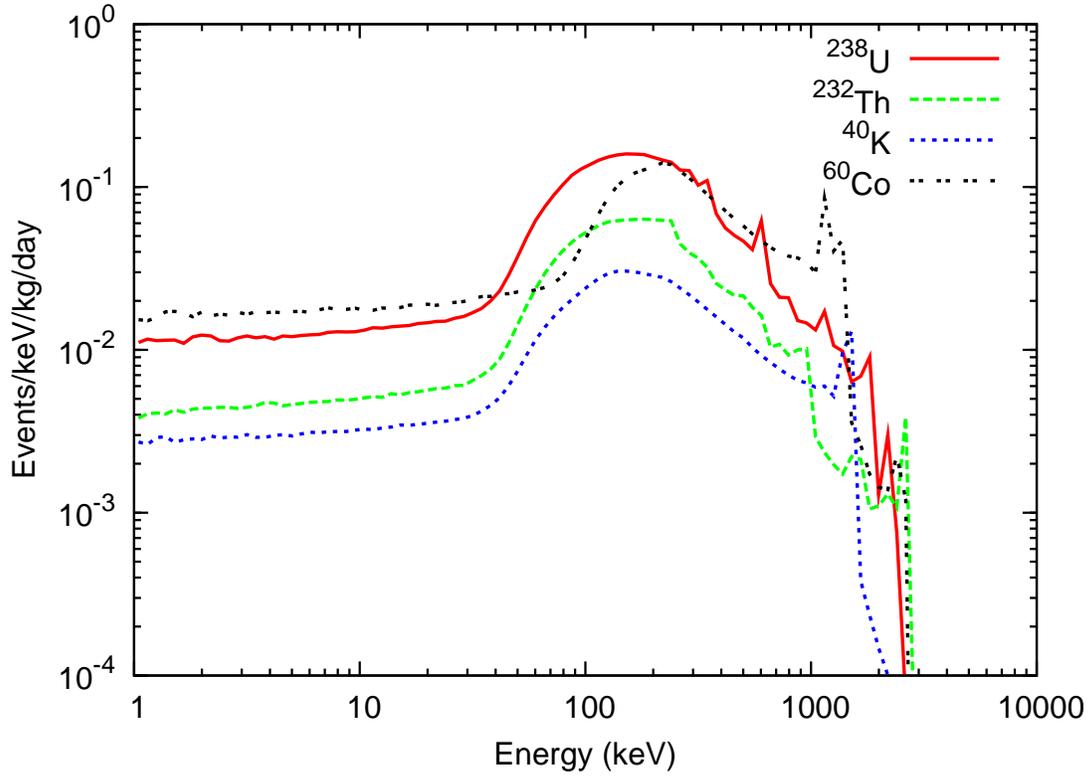}
    \caption{Spectra of energy depositions from electron recoils
in the NaI crystals from $^{238}$U, $^{232}$Th
and $^{40}$K decay chains in secular equilibrium. The source
of radiation was the PMT envelopes (100 g each) attached to the light guides 
connected to the crystals.
Also shown is the spectrum of $^{60}$Co events from Cu. 
Only events in which a single crystal was hit, are included. See text for details.}
  \label{fig:sp-ext}
\end{figure}

Figure \ref{fig:sp-ext} shows the spectra of energy depositions from electron recoils
in the crystal with PMTs contaminated with $^{238}$U, $^{232}$Th and $^{40}$K. 
For normalisation we have used typical concentrations of radioactive isotopes
in ultra-low background PMTs produced by ETL and used by DAMA: 
30 ppb of uranium and thorium, 60 ppm of potassium.
Events produced by $^{60}$Co decays in copper are also shown in Figure \ref{fig:sp-ext}
assuming a decay rate of $^{60}$Co of 10 mBq/kg. This rate is significantly higher
than typically measured in low-background copper.
Only events in which an energy deposition above 0.5 keV is detected
in a single crystal (close to the energy threshold of the DAMA experiment), were included. 
Multiple scattering events in two or more crystals with energy depositions exceeding
0.5 keV were excluded. If a photon scattered two or more times in one crystal, all energy
depositions were summed together. 
The energy deposition spectra were convolved with Gaussian distributions
describing the energy resolution of the DAMA/LIBRA detectors \cite{dama4}.

As can be seen from Figure \ref{fig:sp-ext} the energy deposition spectra from 
all decay chains at low energies from external sources are essentially flat,
due to Compton scattering of high-energy gamma-rays in the crystal.
Spectra from 
any other decay outside the crystal have very similar shape at low energies
dominated by Compton scattered electrons. The assumption
of secular equilibrium does not affect the shape of the spectrum at low energies:
the spectrum from any particular decay has the same flat shape below 30 keV.
The broad peaks at about 150 keV are the well-known back-scatter peaks
due to the scattering of photons in materials around the crystal prior to
entering the sensitive volume.

\subsection{Internal background sources}

\begin{figure}[htb]
   \includegraphics[width=15.cm]{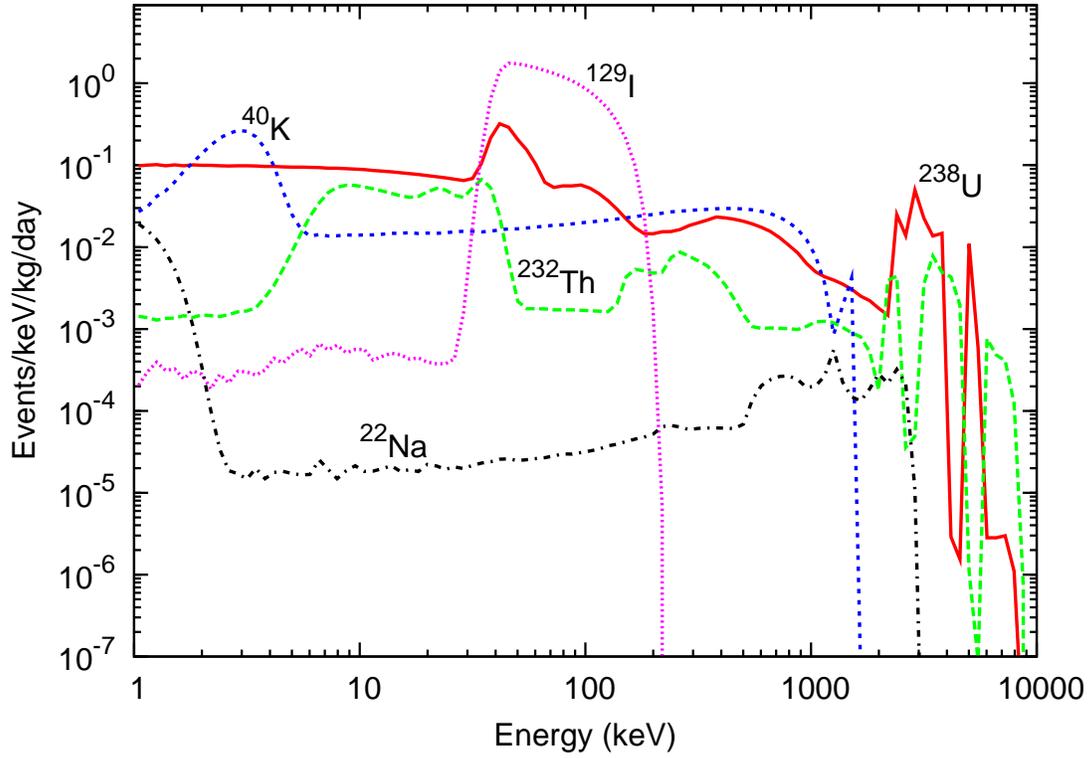}
    \caption{Spectra of energy depositions 
in the NaI crystals from $^{238}$U (solid curve), $^{232}$Th (dashed curve)
and $^{40}$K (dotted curve) decay chains
in secular equilibrium. The decays of $^{129}$I and $^{22}$Na
were also included.
Only events in which a single crystal was hit, are included. The decays of isotopes
occurred within the sensitive volume of the crystals. Spectra were normalised using
the typical concentrations reported in Ref. \cite{dama4}}
  \label{fig:sp-int}
\end{figure}

For case (ii) the sources of background radiation were distributed evenly in all
the crystals. 
Figure \ref{fig:sp-int} shows the energy deposition spectra
from $^{238}$U, $^{232}$Th and $^{40}$K decay chains, as well as from
the decays of some other isotopes ($^{129}$I and $^{22}$Na)
identified as possible sources of background in the DAMA/LIBRA experiment 
\cite{dama3,dama4}. The spectra
were convolved with the energy resolution function.
For energy depositions from $\alpha$-particles we have assumed a quenching 
factor as measured by DAMA \cite{dama4}.

The calculated spectra were normalised
using the measured concentrations of radioactive isotopes
or their decay rates as reported by DAMA \cite{dama4}. 
The main problem here is that only a range of concentrations in the different crystals
can be found in Ref. \cite{dama4}, together with a value determined for one
crystal as an example. The DAMA
Collaboration reported \cite{dama4} the ranges for decay rates of the
isotopes $^{40}$K, $^{129}$I and $^{22}$Na, as well
as for some isotopes from the $^{238}$U and $^{232}$Th
decay chains. We used the reported typical concentrations of these isotopes in 
the crystals
to normalise the spectra shown in Figure \ref{fig:sp-int}:
5 ppt of $^{238}$U (0.7 to 10 ppt was reported in Ref. \cite{dama4} assuming equilibrium), 
5 ppt of $^{232}$Th (0.5 to 7.5 ppt was reported), 10 ppb of natural potassium
($\le$20 ppb was reported), 0.2 ppt of $^{129}$I (a similar value was reported
but for two crystals only) and $6.46 \times 10^{-14}$ ppb of $^{22}$Na.
The $^{232}$Th decay chain was reported to be in equilibrium \cite{dama4}
unlike $^{238}$U for which the equilibrium was found to be broken: a factor of 5.5 higher
rate was found at the end of the chain indicating higher concentration of $^{210}$Pb
than expected from the equilibrium. To plot
the uranium spectrum in Figure \ref{fig:sp-int} we assumed equilibrium
in the uranium decay chain. Since the spectrum from $^{238}$U chain does not 
have any features
below 30 keV, the assumption of equilibrium should not affect the shape of the
spectrum at low energies. Assuming higher concentration of $^{210}$Pb 
compared to the equilibrium would result in a higher peak at 40-60 keV, 
inconsistent with the measurements.

Present simulations were carried out using the most recent version 9.2 of GEANT4. 
The spectra are slightly different from those obtained with the previous version
9.1p3. We have checked the production of the low-energy X-rays, conversion electrons
and Auger electrons in GEANT4 and have found that, despite significant improvements, 
some low-energy photons and electrons are missing resulting in an energy 
imbalance and spectrum distortions if the radioactive decays happened inside the crystal. 
We have corrected this by requiring the conservation of energy, with a missing energy 
being deposited at the point of the decay. The correction 
becomes significant when a low-energy beta-decay is accompanied by the emission of
low-energy X-rays or electrons that are missing from the GEANT4 library, in particular
for $^{129}$I decay.

For case (iii) (radioactive sources distributed in the surface layer), the spectra
(not shown here) are found to be similar to case (ii) with some enhancement in
the number of events at low energies due to particles escaping from the
crystal. This effect, however, does not significantly change the shape of spectra 
at low energies. The assumption that radioactive sources are located
in a thin layer at the crystal surface requires their concentrations to be very high,
the enhancement over the bulk concentration being equal to the ratio of bulk to
surface volumes. Such high concentration looks unrealistic in a well-controlled
low-background dark matter experiment.

\section{Implications for DAMA results}
\label{results}

\subsection{DAMA spectrum}

In this section we will discuss how the measurements and simulations of
the background radiation may help with interpretation of the effect
reported by DAMA. Figure \ref{fig:sp-dama} shows the energy spectra
reported by the DAMA/NaI (open circles) and DAMA/LIBRA (filled circles)
experiments. The spectra,
as reported \cite{dama2,dama3}, were corrected for all detection and
reconstruction efficiencies. 
The energy spectra from the DAMA/NaI and DAMA/LIBRA experiments are slightly
different. This may be due to different cuts to remove PMT noise and different efficiencies
in the two data sets at small energies where the rate is dominated by the PMT noise. 
The difference looks to be significant in the position of the peak
at about 3 keV that is due to $^{40}$K decay \cite{dama4}. 
The energy spectra measured in other NaI detectors at low energies 
are different from those observed by the DAMA/NaI and DAMA/LIBRA experiments.
The filled triangles show the spectrum of energy depositions as measured in NaI
by the NAIAD experiment \cite{naiad}. The DAMA spectra show the minimum 
at about 2 keV that was not seen in NAIAD. However, the absolute rate of events 
was higher in NAIAD than in DAMA/LIBRA. Event rate of about 2 events/kg/day/keV
above 5 keV (similar to DAMA) was observed in Ref. \cite{gerbier99} but the rate in the 
3 keV peak was higher than in DAMA.

The interaction of WIMPs with the target (Na and I) material is 
one interpretation of the measured annual
modulation effect seen by DAMA. The curve in Figure \ref{fig:sp-dama} 
shows the energy spectrum of events from spin-independent WIMP interactions with 
parameters $M_{WIMP} = 60$~GeV and $\sigma_{SI}=7\times10^{-6}$~pb
in the isothermal halo model. These parameters are
close to the best fit to the measured signal as reported in Ref. \cite{dama2}. 
Filled squares show the
difference between the measured spectrum and the spectrum of the WIMP signal
described above.
So the pure background spectrum should look like this curve.
This `background spectrum' shows a deep minimum
at about 1.5-2.0 keV. The depth and exact position of the minimum 
depends on the dark matter particle and halo model
but the minimum should be present in all models, most of them requiring 
the ratio of non-modulated to modulated parts of the signal to be at least 20-30 
at 2-4 keV.
The rise of the measured event rate below 1.5 keV is most likely due to the PMT noise 
that was not included in our simulations.

The rate at 4-10 keV in the DAMA/LIBRA experiment was reported to be about 
1 event/kg/day/keV \cite{dama3,dama4}. 
The peak at about 3 keV is due to the intrinsic contamination
of the crystals by $^{40}$K \cite{dama4}. 
The reported rate at 20-30 keV is about 0.5 events/kg/day/keV.
It is not clear, however, at what energy and why the event rate drops
from about 1 events/kg/day/keV as observed at 4-10 keV to about
0.5 events/kg/day/keV as observed at 20-30 keV. It is difficult to explain 
such a decrease in rate with our simulations. Note, however, that the DAMA
Collaboration reported the spectra at 20-80 keV from two individual crystals only,
whereas the spectrum below 10 keV is averaged over all 25 crystals. 
If the majority of the crystals show much higher rate at 20-30 keV, then
the rates below 10 keV and above 20 keV might be consistent.

\subsection {Combined background spectra from simulations and comparison 
with measurements}

\begin{figure}[htb]
   \includegraphics[width=15.cm]{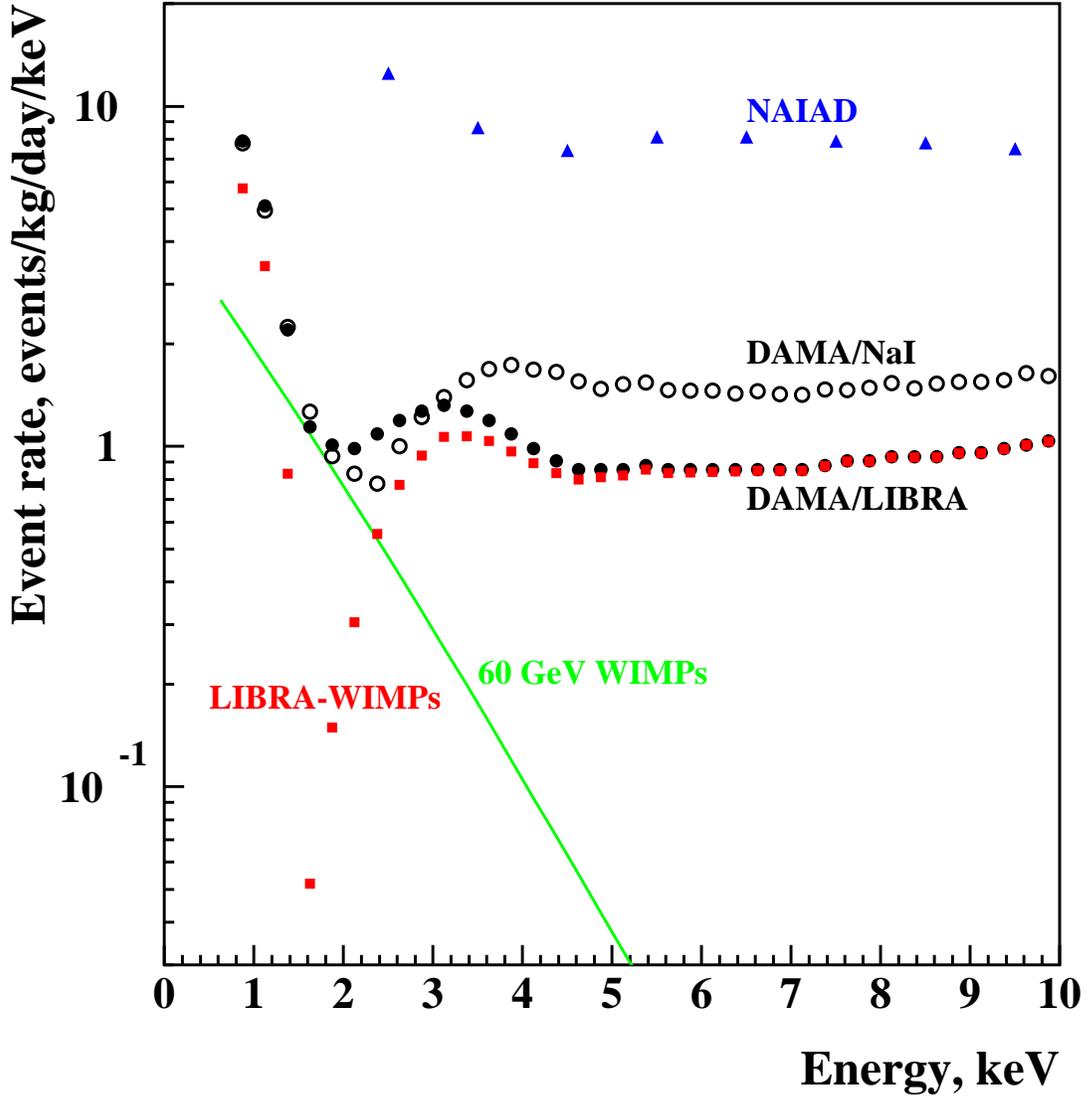}
    \caption{Energy spectra of single hit events as
reported by the DAMA/NaI \cite{dama2} (open circles) 
and DAMA/LIBRA \cite{dama3} (filled circles)
experiments. The spectrum of events expected from 60 GeV WIMP
interactions with the spin-independent cross-section of $7 \times 10^{-6}$ pb
in the isothermal halo model is shown as example by the solid curve (labeled as
`60 GeV WIMPs'). The
difference between the measured DAMA/LIBRA spectrum and the WIMP
signal is plotted as filled squares (labeled as `LIBRA-WIMPs'). 
An example spectrum from one of the NAIAD
crystals is shown by filled triangles.}
  \label{fig:sp-dama}
\end{figure}

\begin{figure}[htb]
   \includegraphics[width=15.cm]{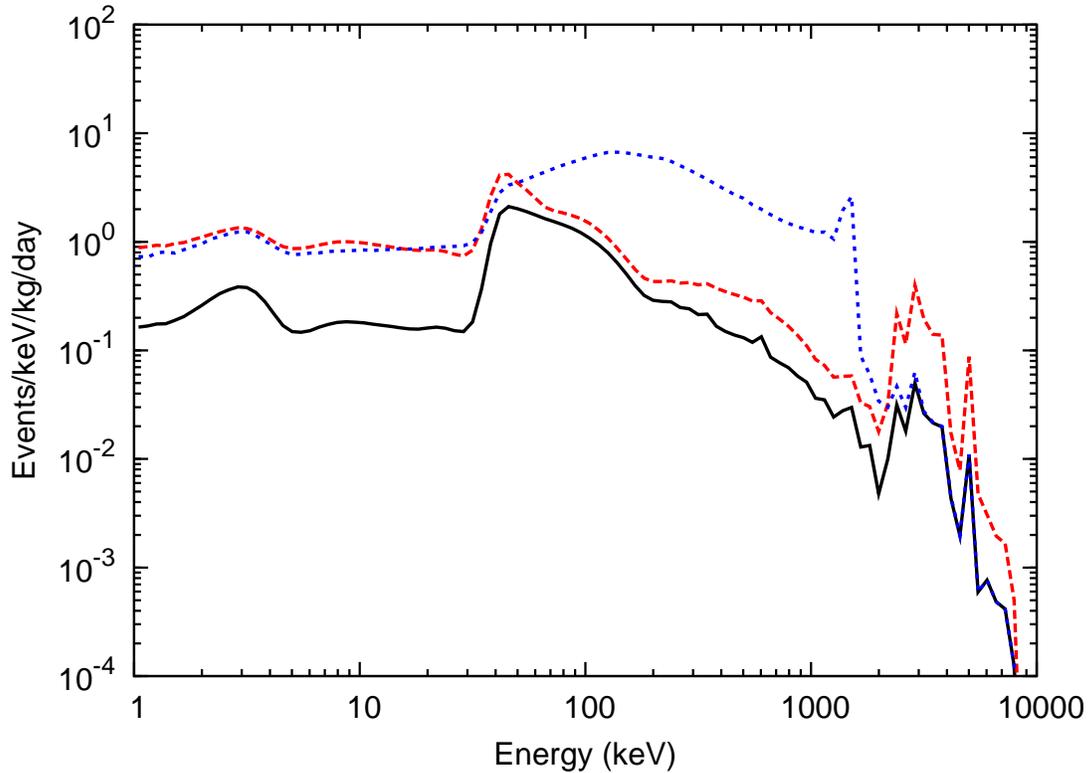}
    \caption{Combined spectra of energy depositions
in the NaI crystals from radioactive background originating in the crystals
and in the PMTs. 
Solid curve: 
PMTs -- 30 ppb of $^{238}$U, 30 ppb of $^{232}$Th, 60 ppm of 
natural potassium in 100 g of the PMT envelope; 
NaI -- 5 ppt of $^{238}$U, 5 ppt of $^{232}$Th, 10 ppb of natural potassium,
0.2 ppt of $^{129}$I and $6.46 \times 10^{-14}$ ppb of $^{22}$Na. 
Dotted curve:  PMTs -- U/Th - the same as for solid curve, 
1.2\% of K; 
NaI -- U/Th - the same as for solid curve, 20 ppt of K.
Dashed curve: PMTs -- the same as for solid curve; NaI -- 
20 ppb of natural potassium, 40 ppt of $^{238}$U, 20 ppt of $^{232}$Th,
other isotopes - the same as for solid curve.}
  \label{fig:sp-sum}
\end{figure}

Figure \ref{fig:sp-sum} (solid curve) shows the sum spectrum of simulated events 
in NaI crystals due to internal and external sources of background as plotted in Figures 
\ref{fig:sp-ext} (except $^{60}$Co) and \ref{fig:sp-int}. The aforementioned concentrations of the
radioactive isotopes were used. Assuming typical concentrations of uranium, thorium
and potassium in PMTs, this external source of background does not
contribute substantially to the total spectrum. The simulated absolute rate below 10 keV
(solid curve in Figure \ref{fig:sp-sum}) 
is significantly smaller than observed by DAMA ($\sim$1 event/kg/day/keV,
see Figure \ref{fig:sp-dama} above and Figure 27 in Ref. \cite{dama4}).
If the copper of the crystal case or shielding, or
other materials are contaminated by $^{60}$Co with a decay rate about
500~mBq/kg, then this source of radiation can explain the rate of events at 4-10 keV
as measured by DAMA.
The broad peak at 40-100 keV is due to the decay of $^{129}$I. The event
rate in the peak is consistent with that reported by the DAMA Collaboration
\cite{dama4}.

In an attempt to make the simulated spectrum compatible with the measurements 
we examined an effect of assuming 
higher concentration of some radioactive isotopes
compared to the reported values. 
The flat part of the spectrum at low energies
could be due to either intrinsic $^{238}$U (with or without equilibrium)
or any external source (decay chains of $^{238}$U or $^{232}$Th, $^{40}$K, $^{60}$Co).
Each spectrum would result in characteristic features at higher energies. 

We have first assumed that the internal contamination of the crystals with radio-isotopes
is higher than reported by DAMA.
The dashed curve in Figure \ref{fig:sp-sum} shows the spectrum calculated assuming
higher concentrations of certain isotopes
inside the crystals compared to the values used in Figure \ref{fig:sp-int}: 
twice as much of natural potassium (20 ppb)
to bring the 3 keV peak to the measured value,
8 times more $^{238}$U (40 ppt) 
to make the spectrum flat at 4-10 keV with a measured
rate of about 1 event/kg/day/keV, and 4 times more $^{232}$Th (20 ppt)
to achieve a better match
at 4-10 keV (see also Figure \ref{fig:sp-int1} below). An important feature
of this spectrum is the presence of a peak from $^{210}$Pb decay at 
$\sim$40-50 keV 
($\beta$-decay accompanied by gamma-rays and X-rays) which dominates
over $^{129}$I decay. 
If the equilibrium is broken in the uranium decay chain and the decay rate of $^{210}$Pb
is higher than that of the parent isotopes as reported by DAMA \cite{dama4}, 
then the peak from $^{210}$Pb will be even
more pronounced making the predicted spectrum inconsistent with the measurements
(see Figure 11 of Ref. \cite{dama4}).

Then we have looked at the possibility of having higher background rate due to external
sources.
Any external source of background radiation would produce a broad 
back-scatter peak centered at about 150 keV. 
The dotted curve in Figure \ref{fig:sp-sum} shows the spectrum calculated
using higher contaminations of potassium in the PMTs (12000 ppm or 1.2\%) 
and in the crystals (20 ppb) 
to account for the flat part of the measured spectrum and absolute rate at
low energies (including the peak at about 3 keV). 
Such a high concentration of potassium (or similarly very
high concentrations of uranium or thorium) in low background
PMTs looks unrealistic but is used to illustrate the effect of possible external
source of background. In fact, such a source could be much closer to the crystal and
not shielded by any light guides or copper. In this case the concentration needed
to obtain the flat spectrum and the rate consistent with measurements
at low energies might be smaller. For instance, as mentioned above, a high decay rate
(about 0.5~Bq/kg) of $^{60}$Co in copper around the detectors (case and shielding)
could explain the observed rate at 4-10~keV.

An important feature of any spectrum dominated by external source,
is the presence of a broad back-scatter peak at about 150 keV. 
Figure 11 in Ref. \cite{dama4} shows, however, that the measured spectrum at 40-80 keV 
(for two individual crystals) 
can fully be explained by the sum of the $^{129}$I and $^{210}$Pb decays
leaving too little room for the back-scattered gamma-rays.
A more detailed investigation of the background could be performed if the measured
spectrum averaged over all crystals for a broad energy range (from 1 keV to a few MeV) 
were to be published by the DAMA Collaboration.

The minimum at 1.5-2~keV that should be present in the background spectrum 
(Figure \ref{fig:sp-dama})
due to the contribution of the WIMP signal to the observed event rate, is not seen
in the simulated spectra (Figure \ref{fig:sp-sum})
independently of the assumptions made about radioactive
contaminations in different materials in and around the detectors.

Figure \ref{fig:sp-int1} shows simulated spectra below 10 keV with
`optimised' concentrations to match the measured spectrum at low energies: 
external background sources were assumed to
be in the PMTs with concentrations specified above for Figure \ref{fig:sp-ext};
internal sources are as for the dashed curve of Figure \ref{fig:sp-int} -- 
potassium at 20 ppb, $^{129}$I at 0.2 ppt,
$^{238}$U at 40 ppt and $^{232}$Th at 20 ppt. The simulated spectrum is
compared with the measurements without signal subtraction (open squares) and
with signal subtraction. For the latter case we assumed 60 GeV mass WIMPs
with $\sigma_{SI}=7\times10^{-6}$~pb (filled squares) and 
$\sigma_{SI}=2\times10^{-6}$~pb (open circles). A similar simulated spectrum but 
over a wide energy range is shown in Figure \ref{fig:sp-sum} by the dashed curve.
Even for a model with small cross-section,
the measured spectrum minus signal (open circles in Figure \ref{fig:sp-int1}) 
does not match the predictions
at 1.5-2.5 keV. A better match between the simulated
background spectrum and the measurements minus signal would require a very small
ratio of non-modulated to modulated parts of the signal spectrum 
(certainly less than a factor of 10 at 2-4 keV). 

Our `optimised' background spectrum, although being in reasonable agreement with the 
measured one, does not satisfy a crucial requirement in the presence of a signal: the
background event rate should not exceed the measured one in any energy bin since
a contribution from a signal is expected too.
In some energy bins in Figure \ref{fig:sp-int1}, in particular at 2-3 keV, the calculated
background rate exceeds the measured one. This can partly be corrected if we assume
an uncertainty in the energy calibration of DAMA/LIBRA. For instance, shifting the
experimental data by about 0.25 keV towards low energies can improve the agreement 
with simulations and opens the way to seeing what the signal spectrum could look like
after background being subtracted from the measured rate.

Figure \ref{fig:sp-int2} shows the measured DAMA/LIBRA spectrum shifted by 0.25 keV
towards low energies. (Note that this is only an assumption and other ways of
improving the agreement between the measured spectrum and the background
taking into account the uncertainties in energy calibration, resolution, etc are possible.)
Let us now require that the expected background spectrum should be less than the
measured rate in any energy bin.
The red solid curve shows the background spectrum with internal concentrations of
34 ppt of $^{238}$U, 20 ppt of $^{232}$Th and 20 ppb of natural potassium.
The reduction of uranium concentration from 40 ppt to 34 ppt is the consequence 
of the above requirement. The background spectrum still looks reasonably close
to the measured one and leaves practically no room for the signal, the difference
between the measured spectrum and the background being shown by filled squares.

To allow for a signal to be present we should now reduce the concentrations of 
radio-isotopes in our calculations. 
Reducing the internal concentration of uranium to 20 ppt 
(dashed blue curve in Figure \ref{fig:sp-int2} corresponds to the sum spectrum) 
results in a 'signal' spectrum shown by open circles which
is obviously inconsistent with any dark matter signal. Internal concentration of
thorium does not affect the spectrum below 6 keV at all (see Figure \ref{fig:sp-int1}).
Lowering the internal concentration of potassium to 10 ppb reduces the
background event rate at the peak only and does not affect energies above 4.5 keV.
The difference between the measured and expected background spectra that can be
associated with the signal, looks like a peak at about 3 keV (filled triangles) 
-- copying the peak from
$^{40}$K. If realistic dark matter models exist with such a feature and with a low ratio
(about 10) of non-modulated to modulated parts of the signal spectrum, they
should be carefully analysed. 
There may be, however, a simpler explanation
of the annual modulation around 3 keV peak related to the selection of single hit 
events at these energies which are dominated by $^{40}$K 
(see also Ref. \cite{gerbier-susy08} for discussion). This can be checked by
looking at the time behaviour of the rate around 1460 keV and of the coincident
events with a 3 keV energy deposition in one crystal and 1460 keV deposition
in another one. 
The problem with a smaller measured rate 
at 40-60 keV compared to the required one from a 
higher concentration of $^{238}$U is still present
in all these scenarios.

If we do not assume high concentrations of some isotopes as suggested above, the
simulated event rate at low energies is much smaller than the DAMA measurements and
no dark matter signal can explain the difference above 6 keV.
At present it is difficult to explain the apparent inconsistency between 
the low-energy part (requiring high concentrations of uranium and 
possibly thorium in the crystals or high rate from an external source) 
and the high-energy part (not showing associated features) of the DAMA spectrum.

Several fits of the measured modulated part of the signal to the predictions of 
different dark matter models were shown in Ref. \cite{dama3}. Unfortunately, no
comparison between the measured spectrum and the predictions including
the background and the time-averaged signal was presented, although such a 
comparison would provide a very important test of the interpretation of
the observed effect.
For instance, most WIMP models that fit 
the DAMA/NaI data \cite{dama2,fs08,sggf08} (some of them are
already excluded by other experiments) are 
not compatible with
the combination of background and signal spectra as shown in the figures.
Most models require the non-modulated part of the signal to be at least 20-30
times bigger than the modulated part at 2-4 keV
resulting in a deep minimum after the signal spectrum
is subtracted from the measured one (see Figure \ref{fig:sp-int1}).
Independently of the relative contributions of external sources
of different isotopes (no accurate measurements for the DAMA experiment were found), 
and even varying intrinsic contaminations of the crystals,
it is hard to explain such a minimum in the 
overall background spectrum at 1.5-2 keV, the measured rate in the peak at 3 keV 
and almost flat rate of 1 event/kg/day/keV at 4-10 keV as
seen in Figure \ref{fig:sp-dama}, with our simulations.

Accurate comparison of the simulations with the measured 
background spectrum in the DAMA crystals
is difficult due to the absence of published data on the spectrum
at all energies (from 1 keV to a few MeV). 
We stress, however, that any analysis of the DAMA signal should 
include predictions of the background spectrum and its comparison
with the measured spectrum minus the averaged dark matter signal extracted from
the annual modulation effect. 

Our results are practically independent of the assumption about the channeling 
effect in NaI crystals (see \cite{dama3} and references therein). Our conclusions
assume that dark matter models predict a high ratio of the non-modulated
to modulated parts of the spectrum (only a small fraction of spectrum
is modulated) which does not depend much on the channeling effect.

\section {Conclusions}
\label{conclusions}
We have presented Monte Carlo simulations of the
background rates and spectra due to different sources expected in the DAMA/LIBRA 
NaI crystals. Any specific source of background radiation
has a characteristic spectrum and, ideally, the measured spectrum
should be consistent with a combination of predicted spectra from the different
sources. We have found that it is hard to match the measured spectrum of 
DAMA/LIBRA at low energies 
(even without signal) with the simulated background assuming measured 
concentrations of radioactive isotopes in NaI crystals. Some features 
(drop in observed intensity between 10 and 20 keV, low total rate from simulations) 
are difficult to explain and 
publication of the measured energy spectrum up to MeV energies by DAMA
is necessary to clarify this point.
If a signal is present, then the background (measured spectrum minus signal) 
should have a minimum at about 1.5-2 keV. Its position and depth depend on the model but 
it is very hard to obtain such a minimum with simulations.
Significantly enhanced concentrations of certain isotopes in the crystals 
help to bring the predicted rate and shape of the spectrum to agreement with
the measured spectrum at low energies but leave too little room for the dark matter signal.
Any dark mater model that predicts a large ratio (above 10 at 2-4 keV) of
non-modulated (average)
to modulated parts of the spectrum can certainly be excluded.
The assumption of higher concentrations of certain isotopes, although improving
agreement with the low-energy part of the spectrum, results in some features that
are not seen by DAMA at high energies.
We do emphasise that any interpretation of the annual modulation of the event rate
observed by DAMA, as dark matter signal, is incomplete without
clear understanding of the background spectrum. Such integrated analysis 
would significantly restrict the 
dark matter models capable of explaining the modulation effect.

\section{Acknowledgments}
This work has been supported by the ILIAS integrating activity
(Contract No. RII3-CT-2004-506222) as part of the EU FP6 programme 
in Astroparticle Physics. We acknowledge the financial support from 
the Science and Technology Facility Council -- STFC).
We thank Prof. G. Gerbier and Dr. R. Lemrani (CEA-Saclay, France) for
fruitful discussions and Dr. L. Pandola (LNGS-INFN) for helping us with generating
radioactive decays with GEANT4. We are grateful to anonymous referees
for useful comments.

\pagebreak

\pagebreak

\pagebreak

\pagebreak

\pagebreak

\begin{figure}[htb]
   \includegraphics[width=15.cm]{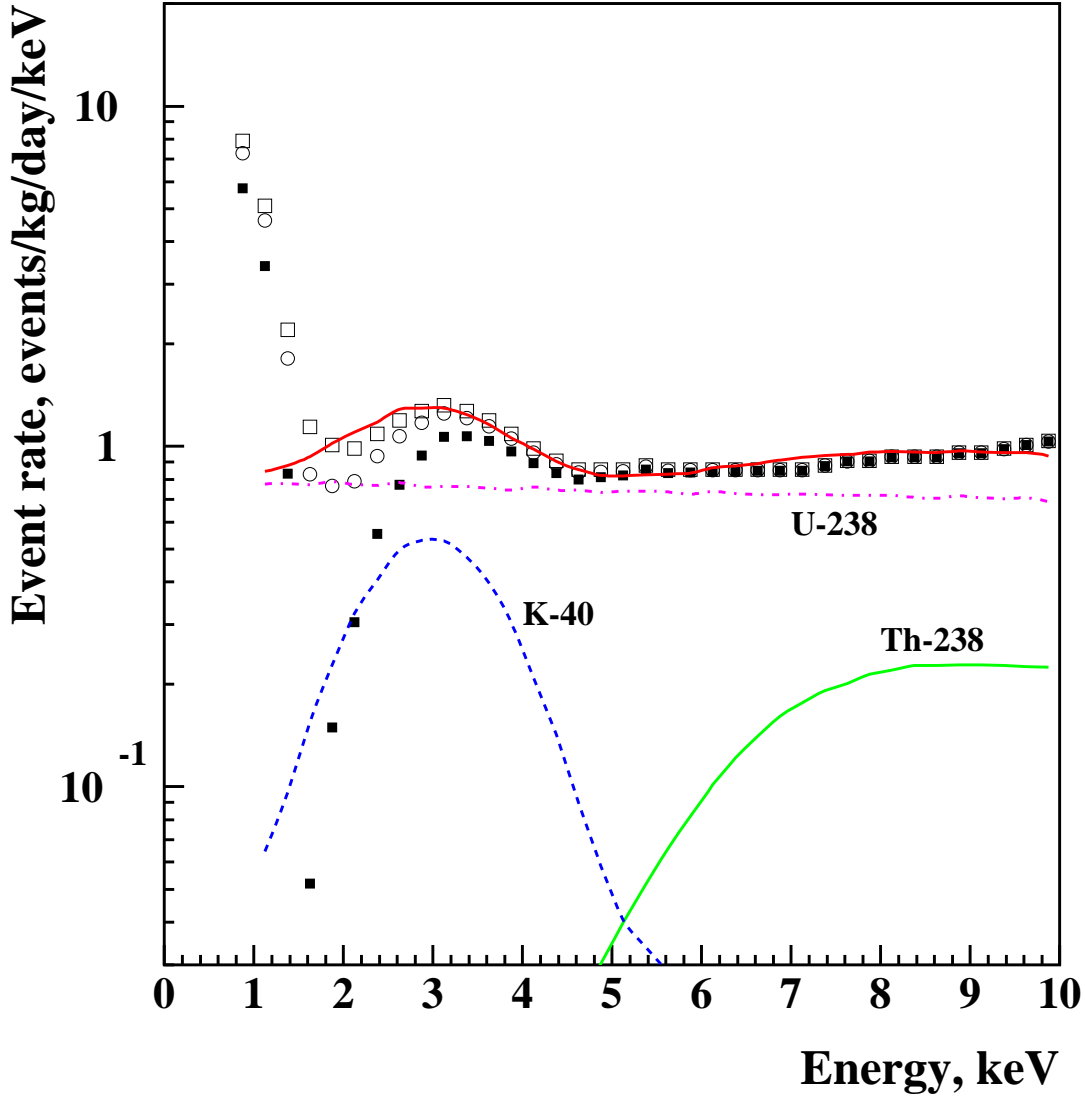}
    \caption{Simulated spectra of energy depositions 
in the NaI crystals from internal sources at low energies
assuming `optimised' concentrations of isotopes in the crystals (solid curve): 
40 ppt of $^{238}$U (pink dashed-dotted curve), 
20 ppt of $^{232}$Th (green solid curve), 20 ppb of natural potassium
(blue dashed curve),
0.2 ppt of $^{129}$I (not seen on the graph). The measured spectrum of DAMA/LIBRA
without signal subtraction (open squares) and
with signal subtraction are also shown. For the latter case we assumed a signal
from 60 GeV mass WIMPs
with $\sigma_{SI}=7\times10^{-6}$~pb (filled squares) and 
$\sigma_{SI}=2\times10^{-6}$~pb (open circles) (see text for details).}
  \label{fig:sp-int1}
\end{figure}

\pagebreak

\begin{figure}[htb]
   \includegraphics[width=15.cm]{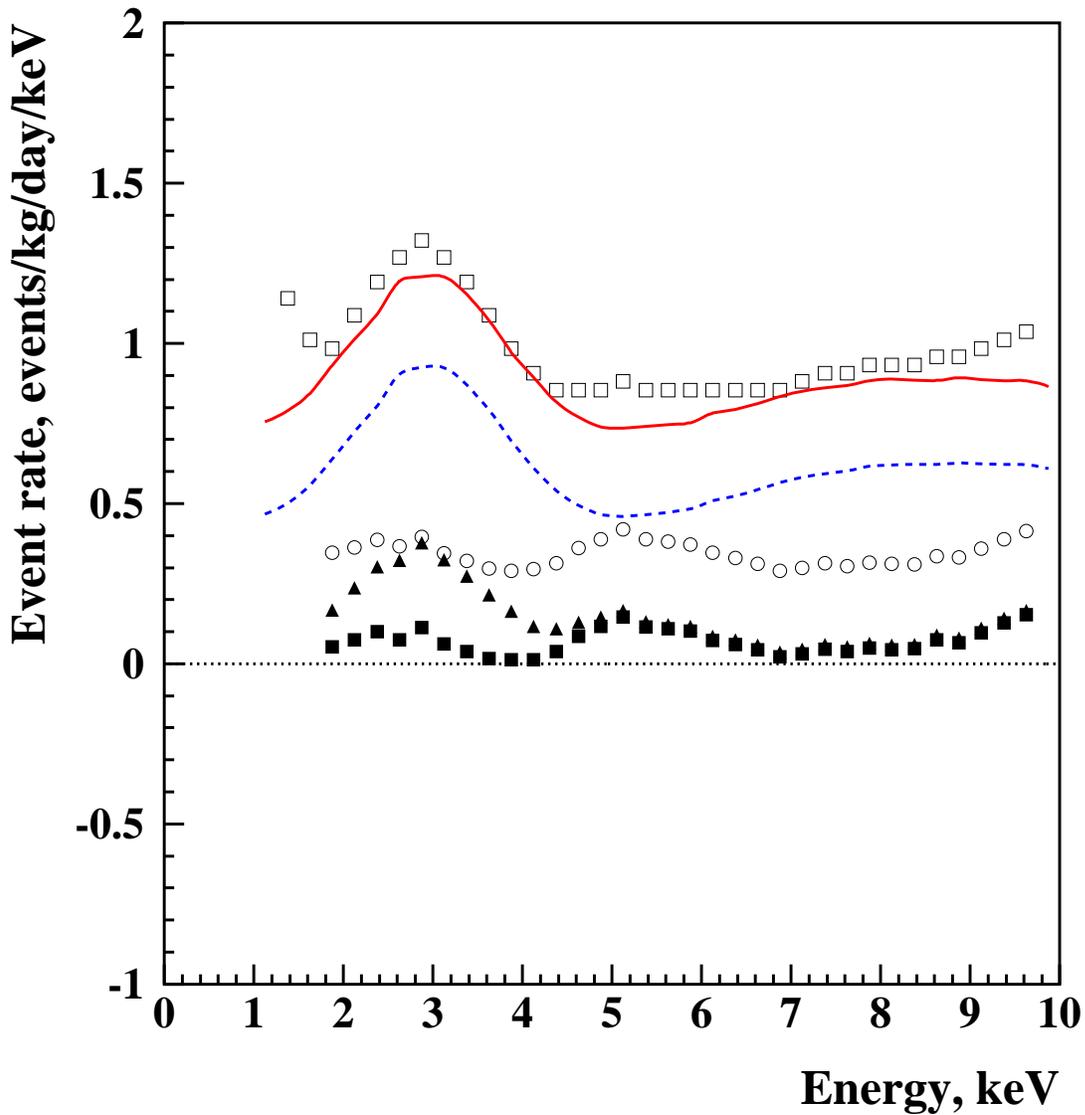}
    \caption{Simulated spectra of energy depositions 
in the NaI crystals from internal sources at low energies
assuming `optimised' concentrations of isotopes in the crystals (solid curve): 
34 ppt of $^{238}$U, 20 ppt of $^{232}$Th and 20 ppb of natural potassium. 
The measured spectrum of DAMA/LIBRA (open squares) is shifted by 0.25
keV towards low energies to achieve better agreement with simulations around
the 3 keV peak. The difference between the two above spectra is shown by filled squares.
The dashed blue line is the calculated spectrum assuming 
20 ppt of $^{238}$U, 20 ppt of $^{232}$Th and 20 ppb of natural potassium
with the corresponding difference between the measurements and calculations
shown by open circles. The difference between the measured spectrum and
the simulated one assuming 34 ppt of $^{238}$U, 20 ppt of $^{232}$Th and 
10 ppb of natural potassium is denoted by filled triangles
(see text for details).}
  \label{fig:sp-int2}
\end{figure}

\end {document}